\RequirePackage{fix-cm}
\documentclass[twocolumn,epjc3]{svjour3}  
\smartqed  
\RequirePackage{graphicx}
\RequirePackage{comment}
\usepackage{color}
\usepackage{lineno}
\usepackage{url}
\usepackage{amsmath}
\usepackage{amsfonts} 

\newcommand{\DBD}{0$\nu$DBD}

\newcommand{\Se}{$^{82}\mathrm{Se}$}
\newcommand{\qvalue}{2997.9$\pm$0.3\,keV}
\newcommand{\duenu}{T$_{1/2}^{2\nu}$=(9.2$\pm$0.7)$\times$10$^{19}$\,y}

\newcommand{\TL}{$^{208}\mathrm{Tl}$}

\newcommand{\CupidZ}{CUPID-0}
\newcommand{\EnrZnSe}{Zn$^{82}$Se}
\newcommand{\dataset}{DataSet}

\newcommand{\exposure}{3.44\,kg$\cdot$y}
\newcommand{\TotalBkg}{(3.6$\pm$0.5)$\times$10$^{-2}$}
\newcommand{\TotEfficiency}{93$\pm$2$\%$}
\newcommand{\CutEfficiency}{95$\pm$2$\%$}
\newcommand{\BetaGammaBkg}{(1.5$\pm$0.3)$\times$10$^{-2}$}
\newcommand{\VetoedBkg}{5.1$^{+2.4}_{-2.0}\times$10$^{-3}$}
\newcommand{\FullVetoedBkg}{3.6$^{+1.9}_{-1.4}\times$10$^{-3}$}

\newcommand{\THO}{$^{232}\mathrm{Th}$}

\newcommand{\ckky}{\un{counts/(keV\cdot kg\cdot y)}}

\providecommand*{\un}[1]{\ensuremath{\mathrm{~#1}}}
\journalname{Eur. Phys. J. C}

\begin{document}

\title{Analysis of cryogenic calorimeters with light and heat read-out for double beta decay searches.}
\author{O.~Azzolini\thanksref{Legnaro}
\and{M.~T.~Barrera\thanksref{Legnaro}
\and J.~W.~Beeman\thanksref{LBNL}
\and F.~Bellini\thanksref{Roma,INFNRoma}
\and M.~Beretta\thanksref{MIB,INFNMiB}
\and M.~Biassoni\thanksref{INFNMiB}
\and E.~Bossio\thanksref{Roma,INFNRoma} 
\and C.~Brofferio\thanksref{MIB,INFNMiB}
\and C.~Bucci\thanksref{LNGS}
\and L.~Canonica\thanksref{LNGS,e2} 
\and S.~Capelli\thanksref{MIB,INFNMiB}
\and L.~Cardani\thanksref{INFNRoma}
\and P.~Carniti\thanksref{MIB,INFNMiB}
\and N.~Casali\thanksref{e1,INFNRoma}
\and L.~Cassina\thanksref{MIB,INFNMiB}
\and M.~Clemenza\thanksref{MIB,INFNMiB}
\and O.~Cremonesi\thanksref{INFNMiB}
\and A.~Cruciani\thanksref{INFNRoma}
\and A.~D'Addabbo\thanksref{LNGS,GSSI}
\and I.~Dafinei\thanksref{INFNRoma}
\and S.~Di~Domizio\thanksref{Genova,INFNGenova}
\and F.~Ferroni\thanksref{Roma,INFNRoma}
\and L.~Gironi\thanksref{MIB,INFNMiB}
\and A.~Giuliani\thanksref{CNRS,DiSAT}
\and P.~Gorla\thanksref{LNGS}
\and C.~Gotti\thanksref{MIB,INFNMiB}
\and G.~Keppel\thanksref{Legnaro}
\and M.~Martinez\thanksref{Roma,INFNRoma,e3} 
\and S.~Morganti\thanksref{INFNRoma}
\and S.~Nagorny\thanksref{LNGS,GSSI,e5} 
\and M.~Nastasi\thanksref{MIB,INFNMiB}
\and S.~Nisi\thanksref{LNGS}
\and C.~Nones\thanksref{CEA}
\and D.~Orlandi\thanksref{LNGS}
\and L.~Pagnanini\thanksref{MIB,INFNMiB}
\and M.~Pallavicini\thanksref{Genova,INFNGenova}
\and V.~Palmieri\thanksref{Legnaro,e4} 
\and L.~Pattavina\thanksref{LNGS,GSSI}} 
\and M.~Pavan\thanksref{MIB,INFNMiB}
\and G.~Pessina\thanksref{INFNMiB}
\and V.~Pettinacci\thanksref{Roma,INFNRoma}
\and S.~Pirro\thanksref{LNGS}
\and S.~Pozzi\thanksref{MIB,INFNMiB}
\and E.~Previtali\thanksref{INFNMiB}
\and A.~Puiu\thanksref{MIB,INFNMiB}
\and C.~Rusconi\thanksref{LNGS,USC} 
\and K.~Sch\"affner\thanksref{GSSI}
\and C.~Tomei\thanksref{INFNRoma}
\and M.~Vignati\thanksref{INFNRoma}
\and A.~Zolotarova\thanksref{CEA} 
}

\institute{
INFN - Laboratori Nazionali di Legnaro, Legnaro (Padova) I-35020 - Italy \label{Legnaro}
\and
Materials Science Division, Lawrence Berkeley National Laboratory, Berkeley, CA 94720 - USA\label{LBNL}
\and
Dipartimento di Fisica, Sapienza Universit\`{a} di Roma, Roma I-00185 - Italy \label{Roma}
\and
INFN - Sezione di Roma, Roma I-00185 - Italy\label{INFNRoma}
\and
Dipartimento di Fisica, Universit\`{a} di Milano-Bicocca, Milano I-20126 - Italy\label{MIB}
\and
INFN - Sezione di Milano Bicocca, Milano I-20126 - Italy\label{INFNMiB}
\and
INFN - Laboratori Nazionali del Gran Sasso, Assergi (L'Aquila) I-67010 - Italy\label{LNGS}
\and
GSSI - Gran Sasso Science Institute, 67100, L'Aquila - Italy\label{GSSI}
\and
Dipartimento di Fisica, Universit\`{a} di Genova, Genova I-16146 - Italy\label{Genova}
\and
INFN - Sezione di Genova, Genova I-16146 - Italy\label{INFNGenova}
\and
CSNSM, Univ. Paris-Sud, CNRS/IN2P3, Universit\'e Paris-Saclay, 91405 Orsay, France\label{CNRS}
\and
DiSAT, Universit\`{a} dell'Insubria, 22100 Como, Italy\label{DiSAT}
\and
IRFU, CEA, Universit\'e Paris-Saclay, F-91191 Gif-sur-Yvette, France\label{CEA}
\and
Department of Physics  and Astronomy, University of South Carolina, Columbia, SC 29208 - USA\label{USC}
}

\thankstext{e1}{e-mail: nicola.casali@roma1.infn.it}
\thankstext{e2}{Present address: Max-Planck-Institut f\"ur Physik, 80805, M\"unchen, Germany}
\thankstext{e3}{Present address: Universidad de Zaragoza, 50009, Zaragoza, Spain}
\thankstext{e5}{Present address: Queen's University, Kingston, K7L 3N6, Ontario, Canada}
\thankstext{e4}{Deceased}


\date{Received: date / Accepted: date}

\maketitle

\begin{abstract}
The suppression of spurious events in the region of interest for neutrinoless double beta decay will play a major role in next generation experiments. 
The background of detectors based on the technology of cryogenic calorimeters is expected to be dominated by $\alpha$ particles, that could be disentangled from double beta decay signals  
by exploiting the difference in the emission of the scintillation light. \CupidZ, an array of enriched \EnrZnSe\ scintillating calorimeters, is the first large mass demonstrator of this technology.
The detector started data-taking in 2017 at the Laboratori Nazionali del Gran Sasso with the aim of proving that dual read-out of light and heat allows for an efficient suppression of the $\alpha$ background. In this paper we describe the software tools  we developed for the analysis of scintillating calorimeters and we demonstrate that this technology allows to reach an unprecedented background for cryogenic calorimeters.

\keywords{Double beta decay \and bolometers \and scintillation detector \and isotope enrichment}
\end{abstract}

\section{Introduction}
\label{intro}
As of today, we do not know any process in nature that violates the total number of leptons L or the number of baryons B, even if the Standard Model of Particle Physics does not predict the conservation of such quantities. On the contrary, the Standard Model predicts, also non-perturbatively,  the conservation of a simple combinations of these numbers: B-L~\cite{Dell'Oro:2016dbc}. A violation of this quantity would be a clear hint of physics beyond the Standard Model and this is one of the reasons motivating the endeavor to search for a never-observed process: the neutrino-less double beta decay (\DBD)~\cite{Furry,Feruglio:2002af}.
This process is a hypothesized nuclear transition in which a nucleus decays with no neutrino emission: $(A,Z) \rightarrow(A,Z+2)+2e^{-}$. 

The importance of \DBD\ resides also in the fact that it can occur only if neutrinos coincide with their anti-particles, so its detection would allow to establish the ultimate nature of this elusive particle.
Finally, the measurement of the \DBD\ half-life T$_{1/2}^{0\nu}$ would provide some insight into the absolute mass of neutrinos~\cite{Strumia:2005tc}.


\section{Scintillating Cryogenic Calorimeters}
\label{sec:bolscint}
The analysis techniques described in this paper apply to experiments using the technology of cryogenic calorimeters (historically also called bolometers). 
A cryogenic calorimeter is made by a temperature sensor coupled to a crystal, which acts as energy absorber. The interactions in the crystal release an amount of energy that gives rise to a sizable temperature variation ($\Delta T \propto \Delta E/C$), provided that the crystal thermal capacity $C$ is low enough. To this aim, the crystals are cooled at cryogenic temperatures (about 10 mK).
The main advantages of this technique, originally proposed by Fiorini and Niinikoski \cite{Fiorini:1983yj}, are the energy resolution (as good as 0.1$\%$) and an efficiency on \DBD\ larger than 80$\%$. Furthermore, the crystals can be grown with high intrinsic radio-purity starting from most of the emitters of interest for \DBD.

The first tonne-scale experiment based on cryogenic calorimeters is the Cryogenic Underground Observatory for Rare Events (CUORE~\cite{Artusa:2014lgv}), now in data-taking at Laboratori Nazionali del Gran Sasso (LNGS) in Italy. 
The analysis of the first months of data (corresponding to an exposure of 86.3 kg$\cdot$y) proved that the detector can reach the target energy resolution and background, and allowed to place a 90$\%$ C.L. lower limit of T$_{1/2}^{0\nu}$($^{130}$Te) $>$1.3$\times$10$^{25}$\,y alone, and of T$_{1/2}^{0\nu}$($^{130}$Te) $>$1.5$\times$10$^{25}$\,y combined with its ancestors Cuoricino and CUORE-0~\cite{Alduino:2017ehq,Andreotti:2011822,Alfonso:2015wka}.

Today, the CUPID (CUORE Upgrade with Particle IDentification~\cite{Wang:2015taa,Wang:2015raa}) interest group is defining the strategy for a future upgrade of CUORE that will allow to increase the sensitivity on the half-life of \DBD\ above 10$^{27}$\,y~\cite{Artusa:2014wnl,Poda:2017jnl,Bellini,Pirro}.

The main challenge for the CUPID project will be the development of a background-free detector at the tonne-scale level. 
The first important milestone is the abatement of the dominant source of background of CUORE, i.e. $\alpha$ particles produced by the materials constituting the detector structure~\cite{Alduino:2017}. 
It was proved~\cite{Pirro:2005ar} that the $\alpha$ interactions can be rejected by coupling each calorimeter with a second detector, specialized in the measurement of the scintillation light emitted by the interactions in the crystal. Unfortunately, TeO$_2$ does not scintillate at cryogenic temperatures~\cite{Casali:2013bva}.
For this reason, the  LUCIFER~\cite{Beeman:2013sba} and LUMINEU~\cite{Barabash:2014una} collaborations focused on the development of a new class of scintillating crystals based on \DBD\  emitters characterized by a high Q-value.
Indeed, choosing \DBD\ candidates with high Q-value, such as $^{82}$Se or $^{100}$Mo, provides a natural reduction of the background contribution from environmental $\gamma$'s, that drops above the 2.6\,MeV $\gamma$-line of \TL.
An extensive R$\&$D activity allowed to characterize the properties of several compounds grown with such emitters, like ZnSe~\cite{Arnaboldi:2010jx,Beeman:2013vda}, ZnMoO$_4$~\cite{Gironi:2010hs,Beeman:2012ci,Beeman:2012jd,Beeman:2011bg,Cardani:2013mja,Armengaud:2015hda,Berge:2014bsa} or Li$_2$MoO$_4$~\cite{Cardani:2013dia,Bekker:2014tfa,Artusa:2016maw,Armengaud:2017hit,Buse:2018nzg}. 
These R$\&$Ds demonstrated that the simultaneous read-out of light and heat in scintillating calorimeters enables a very effective suppression of the $\alpha$ background. To prove the potential of this technology on a medium-scale experiment, we designed and constructed the CUPID-0 detector~\cite{Azzolini:2018tum}, now in data-taking at LNGS.


\section{The CUPID-0 Detector}
\label{sec:detector}
The \DBD\ emitter chosen by the \CupidZ\ collaboration is \Se. This isotope features a Q-value (\qvalue~\cite{Lincoln:2012fq}) well above the 2615\,keV end-point of the natural radio-activity, and its half-life for the 2$\nu\beta\beta$ mode is long enough (\duenu~\cite{2nuSeHalfLife}) to prevent background from pile-up in a tonne-scale experiment.

The Se powder was enriched from its natural abundance to 95$\%$~\cite{Beeman:2015xjv}, and embedded in 24 \EnrZnSe\ crystals (plus 2 natural ZnSe crystals)~\cite{Dafinei:2017xpc}. 
The total mass of the 24 \EnrZnSe\ crystals amounts to 9.65 kg (5.13\,kg of \Se), while the two natural crystals have a total mass of about 850\,g (42\,g of \Se).
The ZnSe crystals are surrounded by a VIKUITI multi-layer reflecting foil produced by 3M, and arranged in 5 towers using a NOSV copper\footnote{made by Aurubis: https://www.aurubis.com/en} structure and PTFE supports. 
Each ZnSe is interleaved with two light detectors (LD).
These LDs consist of disk-shaped Ge crystals (170\,$\mu$m thick and 4.4\,cm in diameter) similar to those described in Ref.~\cite{Beeman:2013zva}. One of the best ways to obtain high-performance LDs at 10\,mK consists of operating also the LDs themselves as cryogenic calorimeters: photons impinging on the LD increase its temperature and are recorded as thermal pulses.

To convert the energy deposits in ZnSe crystals and LD in readable voltage signals, each crystal was equipped with a Neutron Transmutation Doped (NTD) Ge thermistor~\cite{thermistor} using a semi-automatic gluing system. In addition, a Si Joule heater was attached to both detectors to inject a reference pulse, which allows to correct for thermal drifts~\cite{Arnaboldi:2003yp,Andreotti2012}.

The CUPID-0 detector is hosted in the same $^{3}$He/$^{4}$He dilution refrigerator that was used for the CUORE-0 experiment, after some major upgrades to the electronics and to the vibration reduction system. The reader can find in Ref.~\cite{Azzolini:2018tum,Arnaboldi:2018yp} a more extensive description of the cryogenic facility, electronics and data-acquisition, as well as more details about the detector construction, operation and optimization.

\section{Data Collection and Production}
\label{sec:Collection}
An interaction in the ZnSe crystal results in an amplified signal with amplitude ranging from tens to hundreds of mV, a rise time of 10\,ms and a decay time of 40\,ms. Due to the smaller detector sizes, the LD signals are usually faster, with rise-time of a few ms and decay-time of about 8\,ms.
Each channel can be biased, amplified and filtered using a dedicated read-out chain, which allows to optimize the amplification gain and the cut-off frequency of the anti-aliasing filter~\cite{Carniti2016,arnaboldi20018,arnaboldi2015,arnaboldi2010,PIRRO2006672,Arnaboldi:2006mx,Arnaboldi:2004jj,AProgFE}. 
Due to the slow time-development of the recorded pulses and the low detector rate (2\,mHz in physics runs), the data are digitized with sampling frequencies of 1\,kHz for ZnSe and 2\,kHz for LD, and the continuous data stream is transferred to disks for the off-line analysis. 
The data collection is made with a DAQ software package (``Apollo''~\cite{ThesisDiDomizio,ThesisCopello}) that in the past was used for CUORE-0 and it is now being used by CUORE.

The trigger is software generated, and allows to use different algorithms according to the experimental needs. The data presented in this paper are processed with two triggers. For the ZnSe calorimeters we use a trigger algorithm with a channel-dependent configuration that fires when the signal derivative stays above threshold for a certain amount of time. For the LD we use simultaneously the derivative trigger and a second (off-line) trigger that forces the acquisition of the LD waveforms every time the ZnSe trigger fires.
The implementation of the second trigger was motivated by the fact that most of the energy produced by an interaction is dissipated as heat in the ZnSe, while only a few $\%$ escapes the ZnSe crystal in the form of scintillation light: a $\sim$MeV deposit in the ZnSe crystal corresponds to about 10 -- 100 keV in the LD (depending on the crystal as well as on the nature of the interacting particle).
To prevent the loss of small (or noisy) light signals, when a signal is detected in the ZnSe we also associate to it the corresponding waveforms in the LD.
In the following, we will use only this trigger for the LD. The derivative trigger is still run on the light detectors for future analyses (for example, to study events that are not produced by scintillation of ZnSe).

The complete data-stream of all channels recorded by the DAQ, as well as the trigger positions, are saved in NTuples based on the ROOT software framework. 
At the first stage, we convert the continuous data into acquisition windows of 5\,seconds for the ZnSe crystals (4\,s after the trigger and 1\,s before) and 1\,s for the LD (800\,ms after the trigger and 200\,ms before). 
The pre-trigger window is used to compute the baseline value, and thus the detector gain before the interaction occurred~\cite{alessandrello1998,alfonso2018}.

Other informations to be accessed during the off-line analysis, such as the geometrical configuration of the array, the correspondences between ZnSe and LD, the run type (physics, calibration, test...), possible time intervals that have to be rejected because of known problems (earthquakes, electronics problems, major underground activities) are stored in a PostgreSQL database.

Each physics run lasts about  2\,days, and it is followed by a stop of a couple of hours to allow the liquid helium refill of the cryostat and the subsequent stabilization of the detectors. Approximately every month, we perform a calibration of 4 days with \THO\ sources.
Since the most energetic $\gamma$ line produced by \THO\ (2.6\,MeV) is below the \Se\ Q-value, we exploit also other sources to characterize the energy region of the \DBD. To study the energy dependency of the shape parameters in the region of interest (Sec~\ref{sec:HeatPulses}), we use an Am:Be neutron source, emitting a broad distribution of $\gamma$~rays up to several MeV. This calibration was made every time we modified the working parameters of the detectors, in order to prevent possible changes in the shape-dependency on the energy. In the first year of data taking, we performed three Am:Be neutron source calibrations: one during the detector commissioning, one between the physics runs presented in this paper, and one immediately after (as a cross-check). Furthermore, we validated the \THO\ calibration with a $^{56}$Co source, producing $\gamma$ peaks well above the \Se\ Q-value (Sec~\ref{sec:calib}).

The collection of the initial plus final calibration and all the physics runs in between forms a \dataset. With the exception of the first two \dataset s, devoted to the detector optimization, the percentage of live-time for physics analysis (thus excluding calibrations) exceeds 80$\%$.

\section{Heat Pulses Reconstruction}
\label{sec:HeatPulses}
The conversion of the continuous data stream into NTuples containing all the quantities of interest is performed with a C++ based analysis framework (``DIANA'') originally developed for Cuoricino. 
In this section we summarize the processing stages that allow to derive the parameters of interest of the heat pulses. Most of these analysis techniques are very similar to those developed by the Cuoricino, CUORE-0 and CUORE collaborations and are extensively described in Refs.~\cite{Alduino:2016zrl,OuelletThesis,BryantThesis,CushmanThesis}.

\begin{figure}[!t]
\begin{centering}
\includegraphics[width=4.1cm]{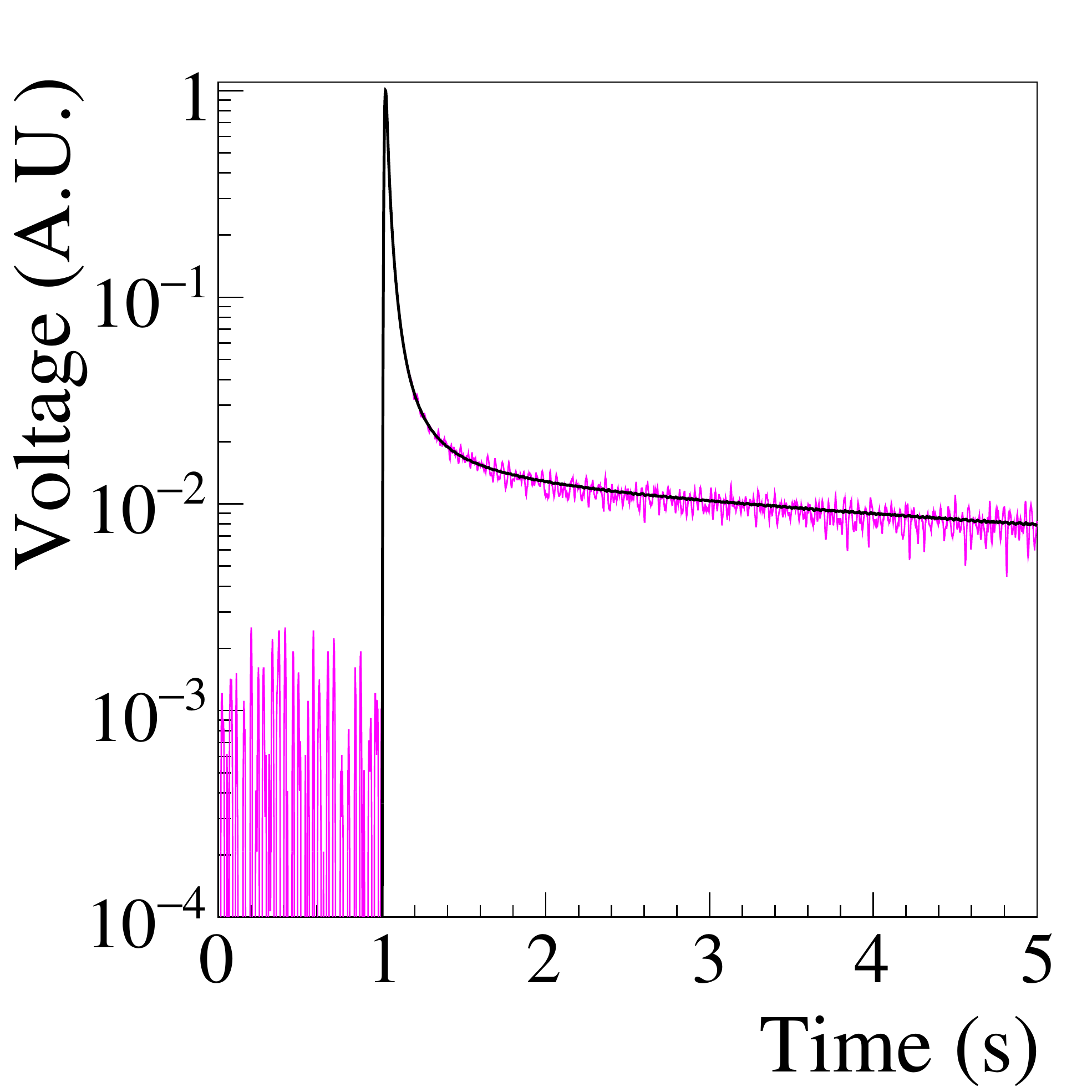}
\includegraphics[width=4.1cm]{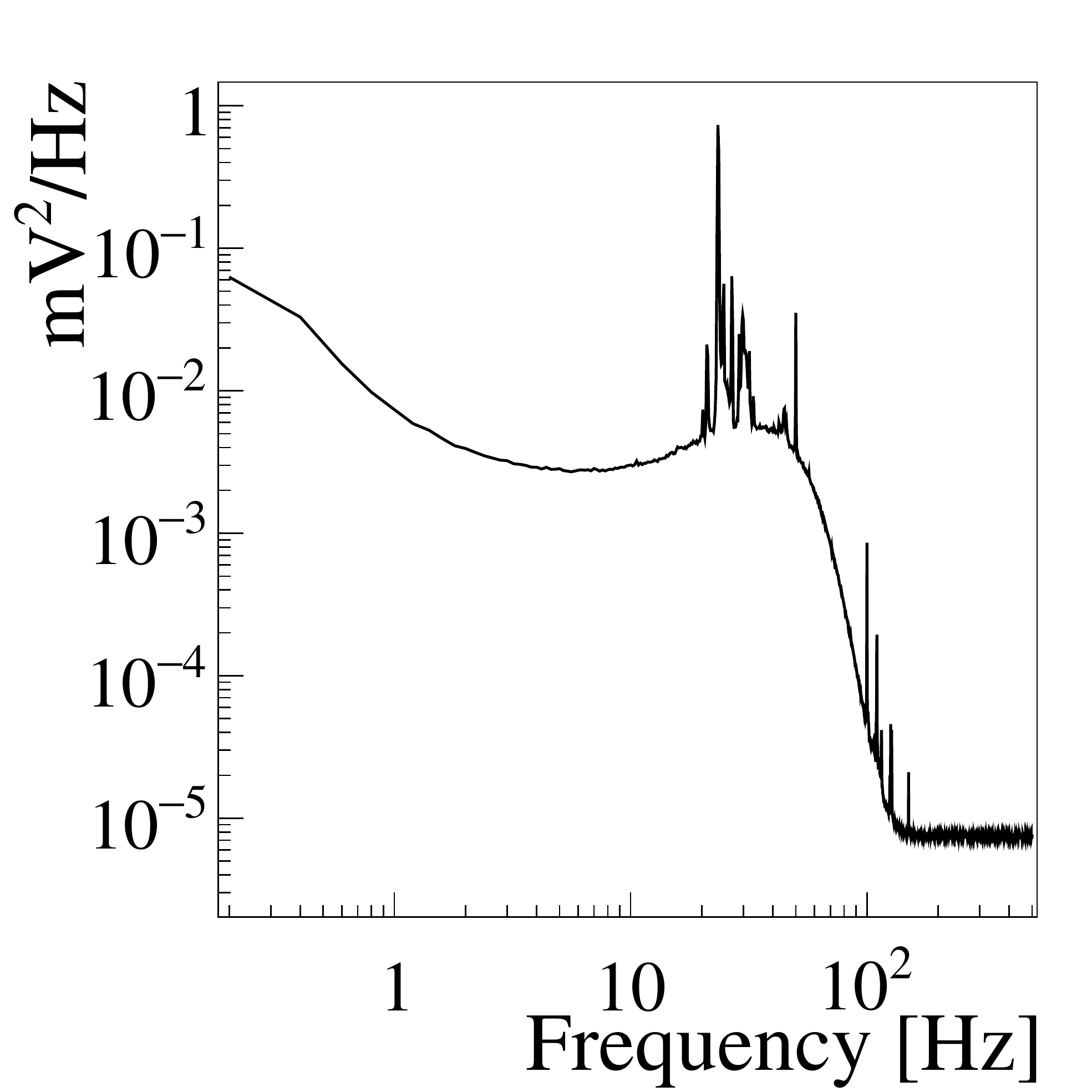}
\caption{Left: a typical template of a ZnSe response (black line) overlapped to a single pulse acquired by the same detector (magenta line). The signal template was evaluated averaging hundreds of pulses in order to suppress the random noise fluctuations. Right: the typical average noise power spectrum of a ZnSe detector. The microphonic noise picks and the roll-off due to the anti-aliasing active Bessel filter are clearly visible.}
\label{fig:PulseNoiseHeat}
\end{centering}
\end{figure}

The heat and light pulses are processed with a matched filter algorithm to suppress the signal frequencies mostly affected by noise and improve the reconstruction of the pulse amplitude~\cite{Gatti:1986cw,Radeka:1966}. 
This software filter requires (for each channel) a template for the detector response and the noise power spectrum, shown in Fig~\ref{fig:PulseNoiseHeat}. The average noise power spectrum is constructed by averaging hundreds of waveforms acquired during the entire \dataset\ with a random trigger. A further off-line analysis allows to discard acquisition windows characterized by the presence of pulses.

The signal template is obtained by averaging hundreds of high amplitude events collected during the \THO\ calibrations, and aligned by the pulse maxima. In such a way, the random noise contributions are suppressed (see Fig~\ref{fig:PulseNoiseHeat} left). In Sec.~\ref{light} we describe how the production of the signal template was improved to match the needs of scintillating crystals. After the matched filter we extract also some parameters related to the pulse shape: rise-time (time difference between the 90$\%$ and the 10$\%$ of the leading edge), decay-time (time difference between the 30$\%$ and 90$\%$ of the trailing edge), slope of the baseline before the pulse, delay of the position of the maximum of the filtered pulse with respect to the maximum of the template, and two shape parameters called Test Value Left (TVL) and Test Value Right (TVR), that correspond to the $\chi^2$ value between the filtered signal template and the filtered pulse computed on the left and right side of the signal maximum, respectively:

\begin{equation}
\begin{split}
TVL &= \frac{1}{A\omega_L}\sqrt{\sum_{i=i_M}^{i_M-\omega_L}(y_i-As_i)^2} \\
TVR &= \frac{1}{A\omega_R}\sqrt{\sum_{i=i_M}^{i_M+\omega_R}(y_i-As_i)^2} 
\end{split}
\end{equation}

where y$_i$ is the pulse, A and i$_M$ its amplitude and maximum position, s$_i$ the ideal signal pulse scaled to unitary amplitude and aligned to y$_i$, $\omega_L$ ($\omega_R$) the left (right) width at half maximum of s$_i$. 

The signal amplitude computed with the matched filter is corrected for temperature instabilities by exploiting the periodic reference pulse injected with the Si heater~\cite{Alduino:2016zrl}. After the correction we expect a residual instability negligible with respect to the noise fluctuations of the detector.
The response of the ZnSe detectors is then equalized by energy-calibrating the $\gamma$ spectrum through the most intense $\gamma$ peaks produced by the \THO\ source between 511\,keV and 2615\,keV.

In the last stage of the analysis, we compute time coincidences between ZnSe crystals. Rejection of coincidences between crystals plays a major role in the background suppression as, from GEANT-4 simulations, we expect  81.0$\pm$0.2$\%$ of the \DBD\ events to be fully contained in a single crystal~\cite{Azzolini:2018dyb}. 
The coincidence window is optimized by selecting events produced by the \THO\ source in which two ZnSe crystals trigger with a total energy of 2615\,keV, and is set to  20\,ms. Given the counting rate of the detectors during the physics runs, we compute the probability of random coincidences among ZnSe crystals to be 1.7$\times$10$^{-6}$.

\begin{figure}[thb]
\begin{centering}
\includegraphics[width=\columnwidth]{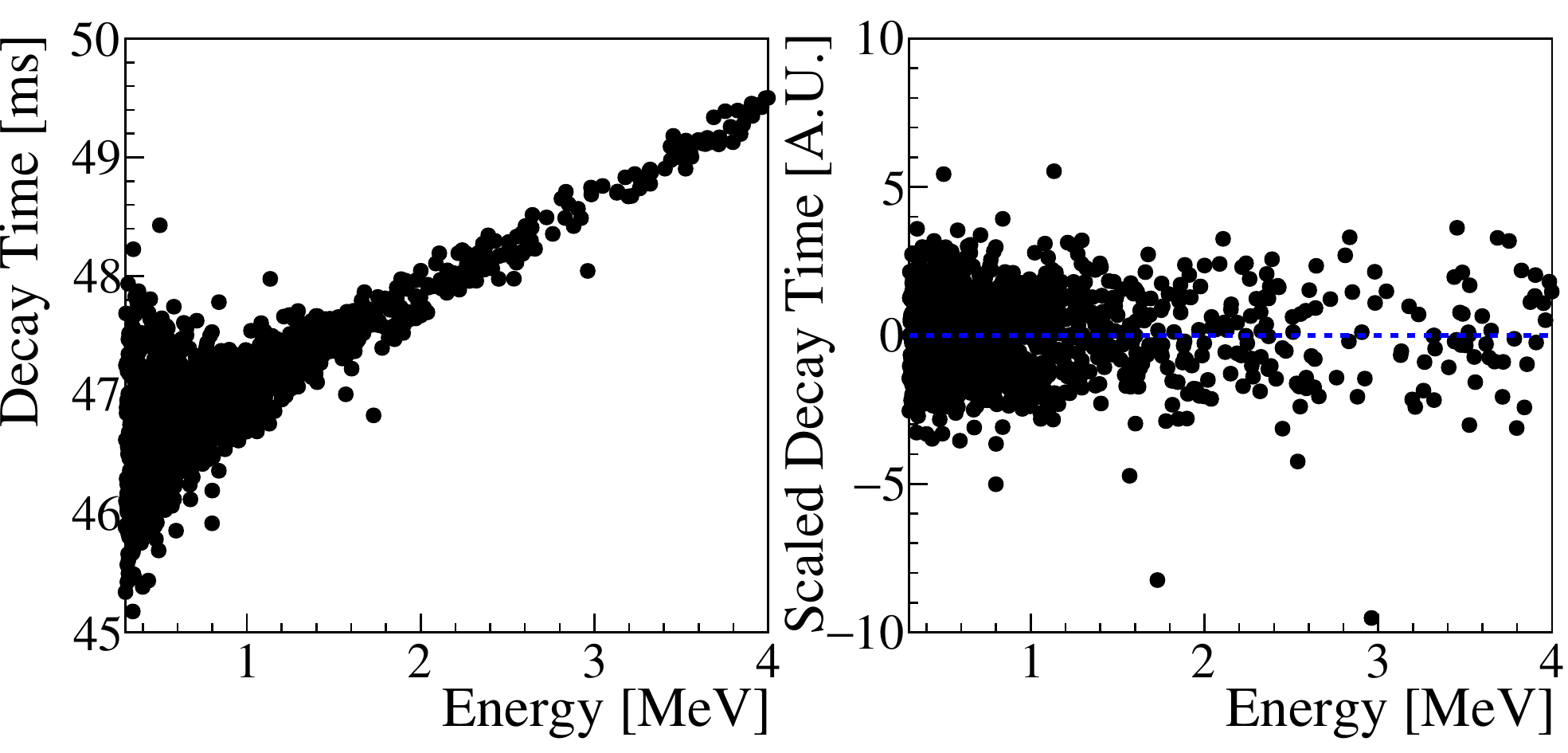}
\caption{Left: decay-time (see definition in text) of pulses recorded by a ZnSe crystal as a function of the energy. Right: same parameter after the removal of the energy-dependency.}
\label{fig:ShapeNormalization}
\end{centering}
\end{figure}

Finally, we estimate and remove the energy dependency of the shape parameters on both their absolute values and resolutions (see Fig.~\ref{fig:ShapeNormalization} left), that otherwise would limit our capability to define robust cuts on the pulse shape. To correct the energy dependency on a wide energy-range, we exploit the periodical calibration with the Am:Be neutron source, producing $\gamma$ interactions (and thus particles with the same shape of the \DBD\ signal) up to several MeV. First of all, for each channel, the energy spectrum is divided in slices, for each of them the median and the MAD (median average deviation) of the considered shape parameter are evaluated. Then, we interpolated both the median and MAD points with a polynomial functions obtaining, in such a way, their trends in the whole energy range (up to 4\,MeV), including the region where we would expect the \DBD\ signal. Finally, we use the parameters extracted from the fits to correct for the energy dependency in all the physics and calibration runs. Each shape parameter is scaled in such a way to be centered around zero (Fig.~\ref{fig:ShapeNormalization}). 

To monitor the effect of this normalization, we developed a software tool which checks that the scaled parameters do not depend on the energy, on the channel and on the measurement time. This analysis led to discard some of the shape parameters (such as the TVR) that are not stable enough to perform uniform cuts over a dataset, as they show a time dependency related to the natural thermal drift of the detectors that can not be corrected.

\section{Selection of Heat Pulses and Efficiency Evaluation}
\label{sec:HeatCut}
We perform a first selection of thermal pulses by exploiting the shape parameters listed in Sec.~\ref{sec:HeatPulses}. 
At this stage, we do not use the information provided by the light detectors, as the purpose is to reject spurious events, such as those barely affected by pile-up or electronics noise.

First, we exclude all the time-intervals that are marked as bad in the database because of known problems (see Sec.~\ref{sec:Collection}). The effect of this selection is a reduction in live-time by 1$\%$. Moreover, we require the pulses to be triggered only by a single ZnSe, as expected from \DBD\ events.

As explained before, we exclude from the analysis the shape parameters that are not robust enough because of fluctuations in time, and we use only the decay-time, rise-time, baseline slope, delay and TVL.

To investigate the effects of cuts on the shape parameters, we study the $\gamma$ peak of $^{65}$Zn, a product of the activation of the Zn contained in the crystals, that decays via electron-capture with a half-life of 224\,d and a Q-value of 1351.9\,keV. This signature acts as a signal sample, while the side-bands close to the $\gamma$ peak are chosen as background samples. 
The odd events are used to optimize the cut while the even ones to compute the selection efficiency.

\begin{figure}[thb]
\begin{centering}
\includegraphics[width=\columnwidth]{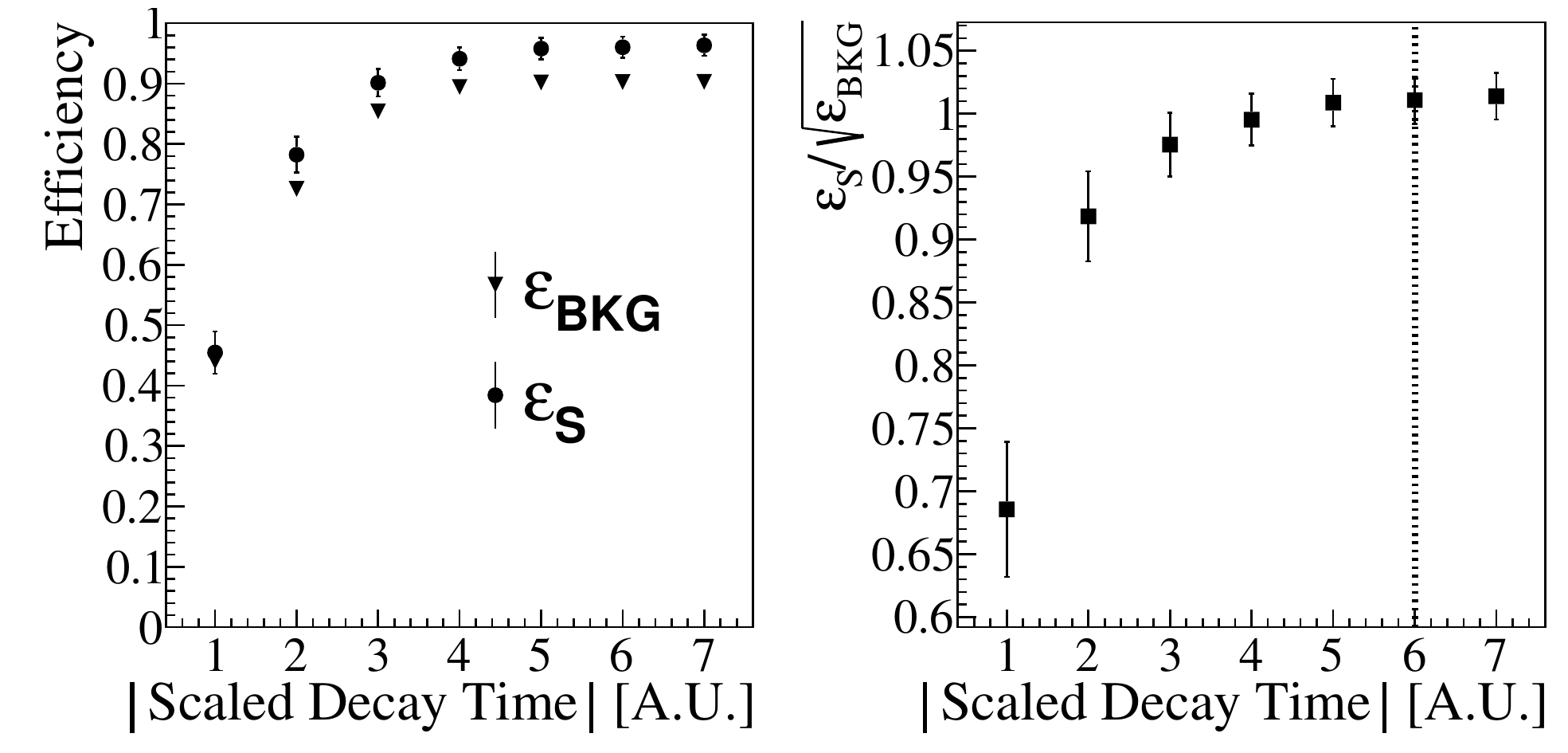}
\caption{Left: efficiency as a function of the integer value at which the scaled shape parameter is cut. Dots: efficiency computed using the $\gamma$ peak of $^{65}$Zn; Triangles: efficiency computed on the side-bands of the peak. Right: ratio $\epsilon_S$/$\sqrt{\epsilon_{BKG}}$; the vertical dotted line represents the chosen cut value. These plots refer to the scaled decay-time, the other parameters show the same behaviour.}
\label{fig:CutOpimization}
\end{centering}
\end{figure}

We scan the distribution of each scaled shape parameter by cutting at different integer values. In Fig.~\ref{fig:CutOpimization}, we report the efficiency on the $\gamma$ peak of $^{65}$Zn ($\epsilon_{S}$) and on its side-bands ($\epsilon_{BKG}$) as a function of the value at which we cut the scaled decay-time. 
This plot shows that $\epsilon_{S}$ is larger than $\epsilon_{BKG}$, proving that the choice of the signal/background samples was reasonable. The reason why they do not differ dramatically, is that the background sample contains also a large fraction of events due to the 2$\nu$ double beta decay that, as expected, is not affected by the shape cuts.
When the cut becomes wide enough, both the efficiencies reach a plateau. To set the proper cut value, i.e. keep the highest efficiency on signal while suppressing the spurious events, we compute the ratio $r =\epsilon_S/\sqrt{\epsilon_{BKG}}$ (Fig.~\ref{fig:CutOpimization}) and choose the cut in which $r$ reaches the plateau.  

\begin{figure}[thb]
\begin{centering}
\includegraphics[width=\columnwidth]{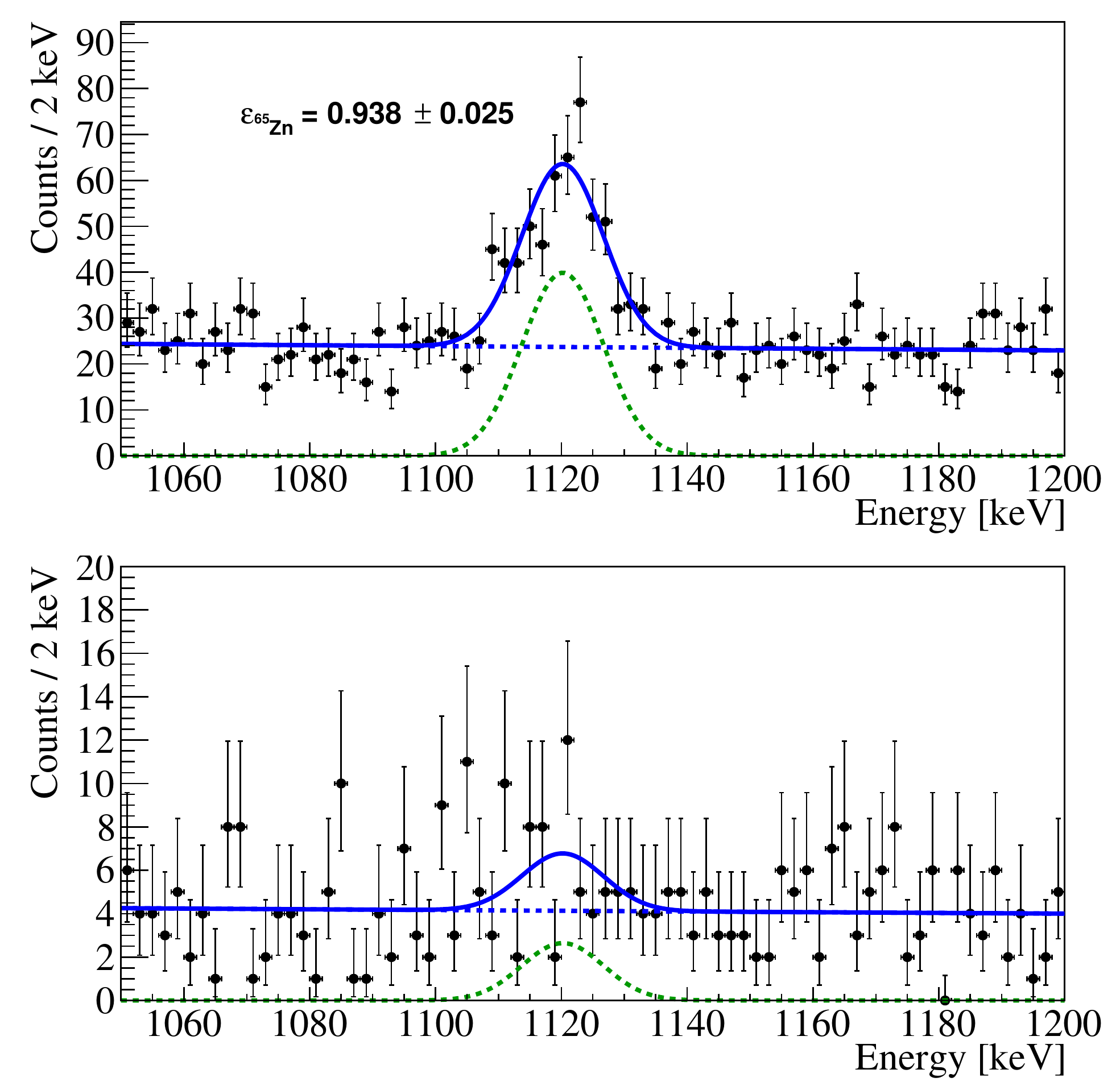}
\caption{$\gamma$ peak of $^{65}$Zn recorded in half a \dataset\ (even events). Top: events that pass the pulse-shape cuts and the anti-coincidence cut. Bottom: events rejected by the pulse-shape cuts. We fit both the plots simultaneously with an unbinned extended maximum likelihood fit with two components: a Gaussian function and an exponential background using the RooFit analysis framework.} 
\label{fig:Zn}
\end{centering}
\end{figure}

As explained before, we evaluate the total efficiency of these cuts on the even events belonging to the $\gamma$ peak of $^{65}$Zn (Fig.~\ref{fig:Zn}). Even if we keep the same shape-cuts during the entire analysis, we compute the efficiency separately on each \dataset\ to account for possible time-variations of the shape parameters.
Weighting the efficiencies by the \dataset\ exposure, we obtain an average efficiency of \CutEfficiency, with a maximum variation of 6$\%$ across all \dataset s.

This value is cross-checked using events in which two crystals triggered that, given the negligible amount of random coincidences, can be considered as an almost pure sample of signal-like events. We obtain an efficiency compatible with the one evaluated on $^{65}$Zn and constant from 300 to 2615\,keV.

The energy region chosen for the analysis of the background is a 400\,keV interval centered around the \Se\ Q-value (2800--3200 keV). At higher energies, indeed, we expect the background to decrease, as the contributions from $^{214}$Bi and $^{208}$Tl (the dominant background sources) drop above 3200\,keV. Therefore, further enlarging the analysis window would result in a lower background. The lowest bound of the interval was chosen to have a symmetric region around the Q-value and, at the same time, to avoid contributions from the 2615\,keV photon or from the tail of the 2$\nu$ double beta decay.

Applying these cuts on the pulse-shape parameters, and requiring that each pulse is triggered by a single ZnSe, we obtain a background index of \TotalBkg\ \ckky.

\section{Validation of the \THO\ Calibration}
\label{sec:calib}
\begin{figure}[thb]
\begin{centering}
\includegraphics[width=\columnwidth]{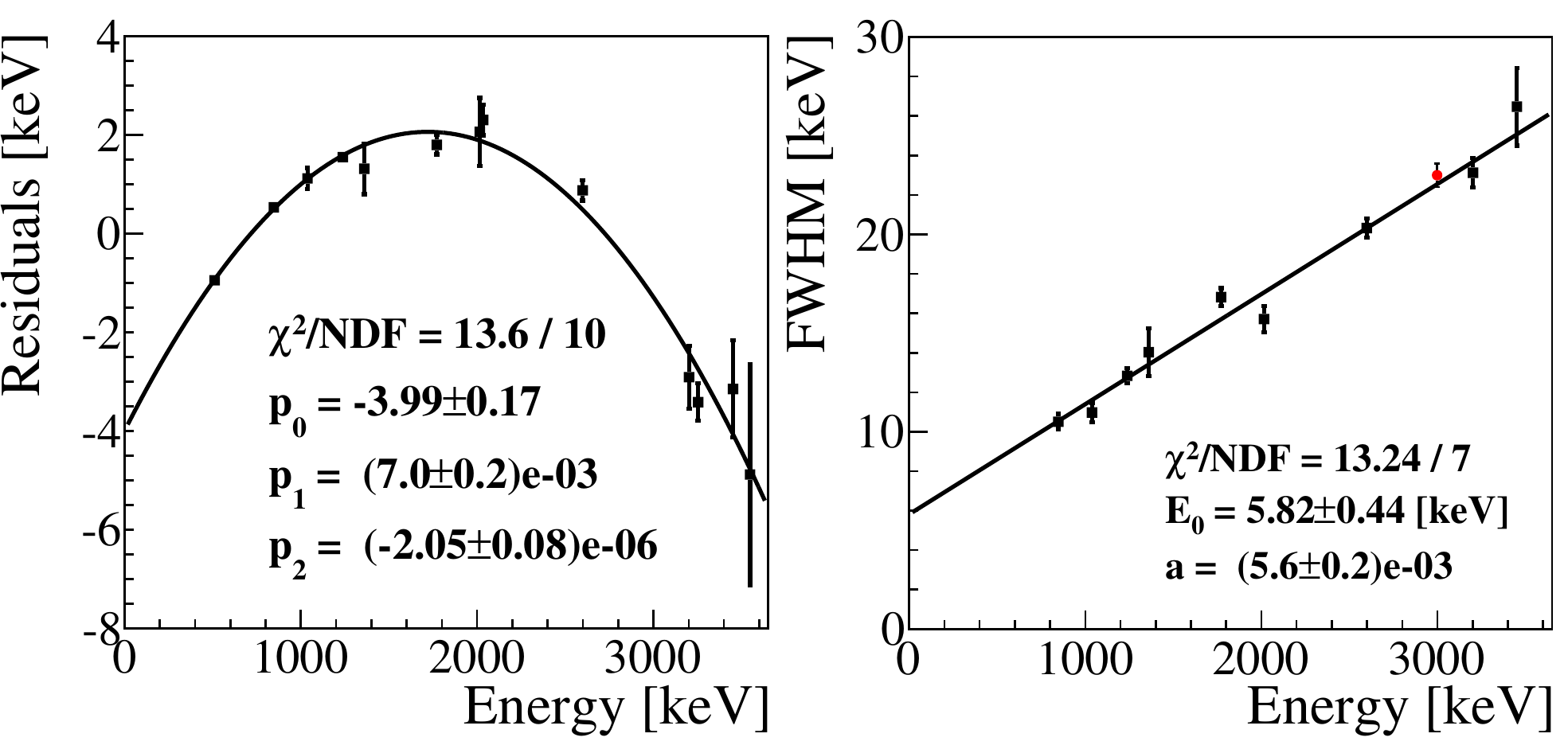}
\caption{Study of the most prominent $\gamma$ peaks of $^{56}$Co, calibrated with the coefficients derived from the \THO\ calibration. Left: difference between the nominal energy and the peak position as a function of the energy. Data are fitted with a parabolic function, resulting in 1.3\,keV residual at the \Se\ Q-value. Right: FWHM energy resolution as a function of the energy (black squares). Data are fitted with a linear function $\Delta E= E_0+aE$. The red circle indicates the value extrapolated from the \THO\ calibration in Ref.~\cite{Azzolini:2018dyb}.}
\label{fig:CoCalibration}
\end{centering}
\end{figure}

Before introducing the information on the light detectors, it is worth observing that the energy calibration of the heat channels was cross-checked in a dedicated measurement.
Indeed, since the Q-value of  \Se\ exceeds the largest $\gamma$ ray produced by the \THO\ source, we usually extrapolate the calibration function and the energy resolution in the region of interest. 
This procedure was used in the analysis of the \DBD\ reported in Ref.~\cite{Azzolini:2018dyb}, in which the extrapolation at high energies resulted in an uncertainty of 3\,keV on the Q-value, and an energy resolution of ($23.0\pm0.6$)\,keV FWHM.
In this paper we validate the \THO\ calibration by using a 17\,days-long measurement with a $^{56}$Co source (T$_{1/2}\sim77.2$\,days) emitting $\gamma$ rays up to $\sim$3.5\,MeV. 
We apply to these runs the calibration coefficients derived for each ZnSe from the \THO\ calibrations, as done in a ``standard'' \dataset. 
We then fit the most prominent $\gamma$ peaks using a double-gaussian model~\cite{Azzolini:2018dyb} and study the difference between the obtained position and their nominal energy, as well as their energy resolution (Fig.~\ref{fig:CoCalibration}).

The residuals show a dependency on the energy that can be modeled with a parabolic function, resulting in a uncertainty on the position \DBD\ peak of 1.3\,keV. This value is negligible compared to the energy resolution in that region, and proves that the choice of an uncertainty of 3\,keV in the analysis of the \DBD\ reported in Ref.~\cite{Azzolini:2018dyb} was very conservative.
The  energy dependency of the energy resolution is modeled with a linear function. In the region of interest we obtain a FWHM of $22.5\pm1.2$\,keV, fully consistent with the value extracted from the \THO\ calibration~\cite{Azzolini:2018dyb}.

\section{Reconstruction of Light Pulses}
\label{light}

\begin{figure}[htb]
\begin{centering}
\includegraphics[width=\columnwidth]{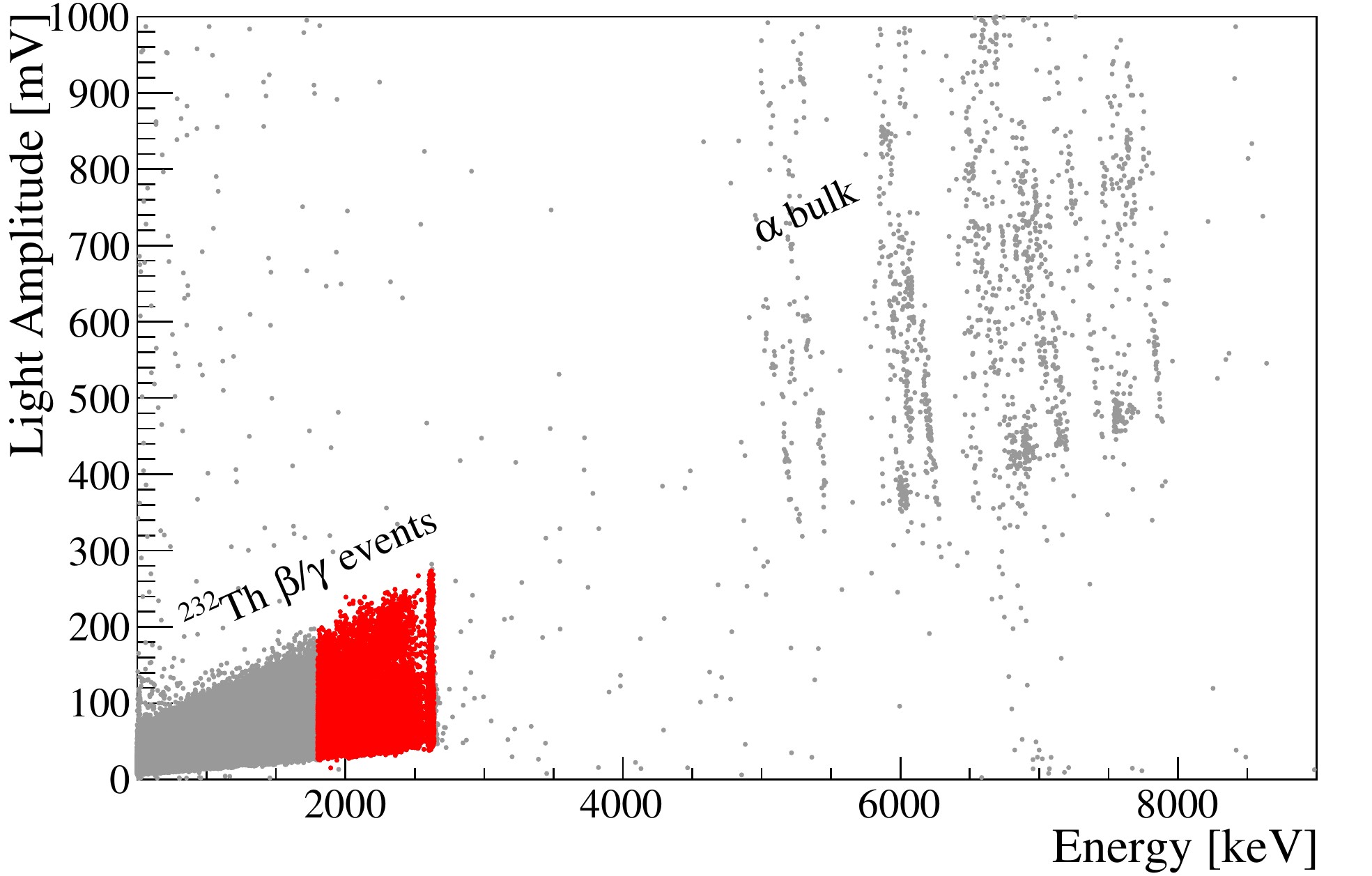}
\caption{The light recorded by the LD placed on top of the ZnSe is reported as a function of the heat released in the crystal. The heat axis is energy-calibrated using the most intense $\gamma$ peaks produced by the \THO\ source. On the contrary, the light axis is not energy-calibrated (broadening the $\beta/\gamma$ and $\alpha$ bands). Red: pulses that pass the selection criteria to produce the signal templates, both for the heat and the light detectors. In this plot, data from all crystals are shown.}
\label{fig:pulses_selection}
\end{centering}
\end{figure}

\begin{figure}[thb]
\begin{centering}
\includegraphics[width=4.1cm]{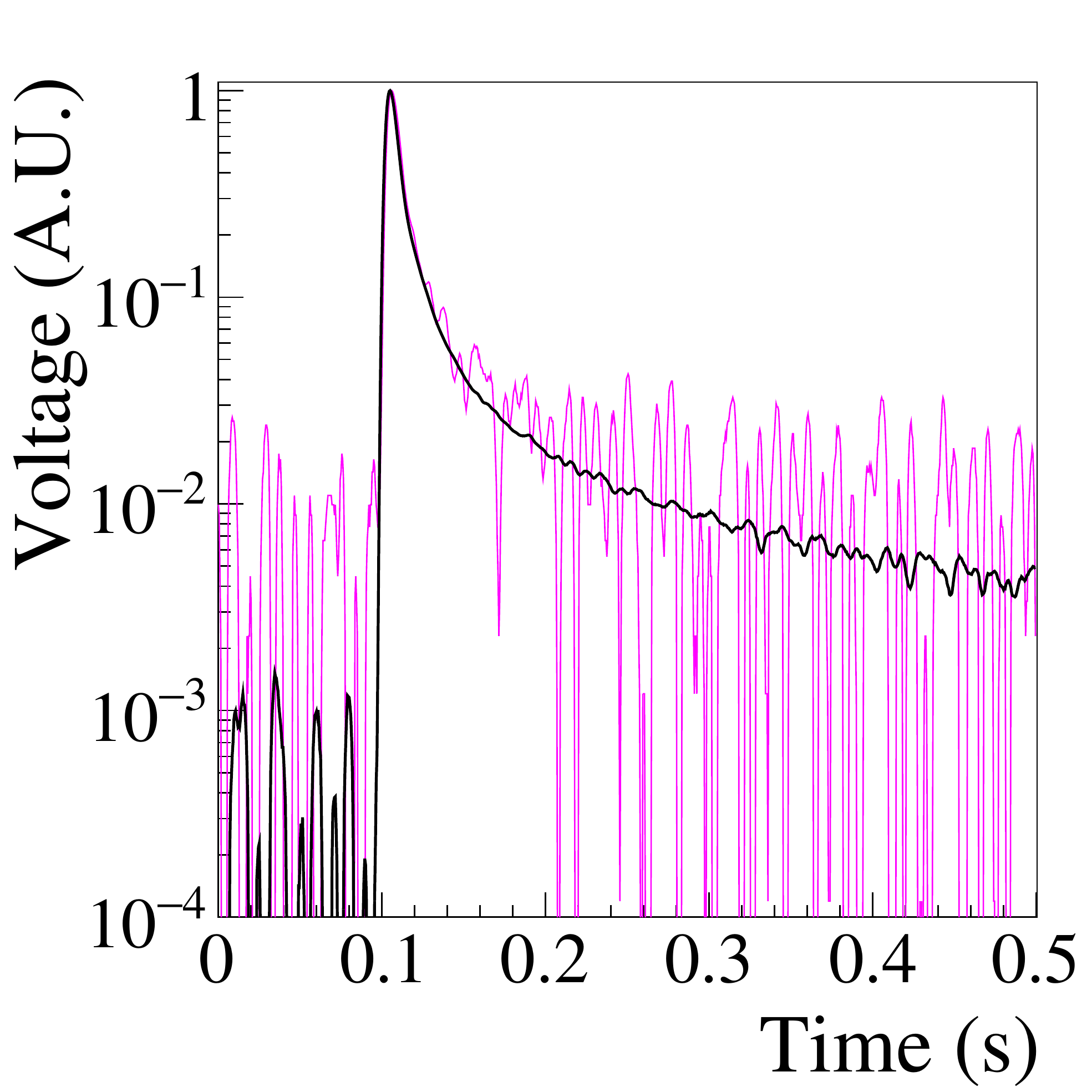}
\includegraphics[width=4.1cm]{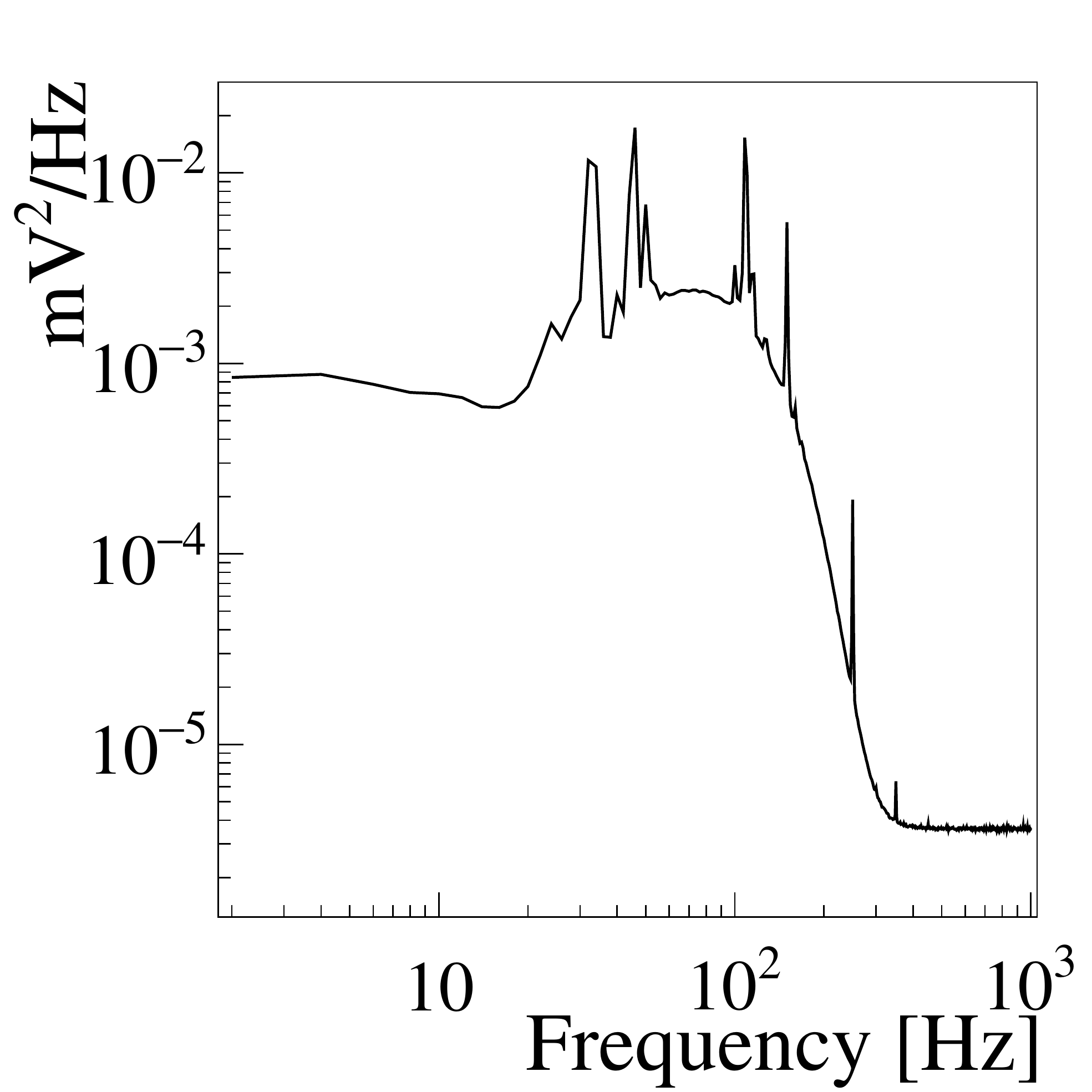}
\caption{Left: a typical template of a LD response (black line) overlapped to a single light pulse acquired by the same detector (magenta line). The signal template was evaluated averaging hundreds of light pulses emitted by the bottom ZnSe crystal in order to suppress the random noise fluctuation. Right: a typical average noise power spectrum of a LD. The microphonic noise picks and the roll-off due to the anti-aliasing active Bessel filter are clearly visible.}
\label{fig:PulseNoiseLight}
\end{centering}
\end{figure}

The first step for a correct reconstruction of the light pulses consists of generating a dedicated signal template for the matched filter.
In the past, the signal response of the LD was made by averaging many pulses with good signal-to-noise ratio, obtained for example using a $^{55}$Fe X-ray source permanently exposed to the detector.
Nevertheless, this is not the best approach, as X-rays, $\alpha$ particles and electrons are characterized by a different time development of the light pulses, and constructing the ideal detector response on a class of events that is not similar to the one of \DBD\ can spoil the evaluation of the light shape parameters. This has a particular importance for \CupidZ\ that, as explained in Sec.~\ref{sec:rejection}, takes advantage from the shape of the light pulses for particle identification. 
For this reason, we developed a new algorithm that selects only events with a shape similar to the one of \DBD. Exploiting the $\gamma$ energy calibration made with the \THO\ source, we select heat pulses with energy of 1.8 -- 2.64 MeV. This energy interval is wide enough to provide a large sample of pulses without introducing non-linearities in the pulse shape. Moreover, we require the emitted light to be compatible with the one produced by scintillation of electrons to discard events produced by scintillation of $\alpha$ particles, or events with no associated light emission (electronics noise, interactions in the NTD Ge sensor). Finally, we reject spurious events, such as those affected by random pile-up, or those in which a second pulse was detected in the same acquisition window. The effect of the selection is shown in Fig.~\ref{fig:pulses_selection}. 
The events that pass all the selection cuts (a few hundreds for each ZnSe) are finally averaged to suppress the random noise contributions. 
For each ZnSe, we construct 3 signal templates: the template of the ZnSe itself, and the templates of the light recorded by the top/bottom LD. We stress that now, in contrast to the past~\cite{Artusa:2016maw}, each LD has two different signal templates, corresponding to the light emitted by the top/bottom ZnSe (see Fig.~\ref{fig:PulseNoiseLight}).
The new structure of the signal templates demands for a new version of the matched filter algorithm with respect to the one described in Ref.~\cite{Alduino:2016zrl}.
From the new matched filter we extract again all the shape parameters of the light pulses. In the analysis presented in this paper we considered only the parameters of the light detectors placed on top of the ZnSe crystals as the SiO coated face is more sensitive with respect to the other one. We observe that the parameter that provides the best particle identification is the TVR (Sec.~\ref{sec:HeatCut}) that in the following will be called $Shape\ Parameter$, or $SP$, for simplicity. A preliminary study indeed, allowed to infer that the background rejection obtained with $SP$ over-performs the one obtained with the light yield alone.

Thanks to the new algorithm used for the light signals analysis, the TVR of the light pulses does not show channel-dependent behaviour. Furthermore, since the LDs are operated at a slightly higher temperature with respect to the ZnSe detectors their working points result very stable over the time. Therefore, the normalization procedure is not needed for this parameter because it turns out to be very stable and reproducible both over the channels and time, as shown in the next section.

Finally, we improve the evaluation of the amplitude of the light pulses that, given the worse signal-to-noise ratio with respect to heat pulses, could be affected by larger uncertainties.
This problem is corrected by measuring the matched-filtered amplitude of the light pulses at a fixed time-delay with respect to the ZnSe scintillation that triggered the event (see Ref.~\cite{Piperno:2011fp}). The main difference with respect to the algorithm described in that paper is the calculation of the time delay. In Ref.~\cite{Piperno:2011fp}, the time-delay was computed as the median of the time intervals between the heat pulses and the corresponding light pulses. In \CupidZ\ we compute the time delay using the filtered signal templates of heat and light. This algorithm does not give a more precise evaluation of the delay, but it is more simple and fast to implement as it requires only the templates of the signals.

In contrast to the ZnSe channels, it is not possible to energy-calibrate the amplitude of the LD using the \THO\ strings placed in the external part of the cryostat. The energy of the scintillation light produced by interactions in the ZnSe crystals usually ranges from a few keV to tens of keV; $\gamma$'s of this energy are too weak to penetrate the external shield of the refrigerator.
In the past, this problem was overcome by depositing an X-ray source on a support permanently exposed to the surface of the LD.  In \CupidZ, we decided to avoid the presence of sources to be conservative from the point of view of the radioactivity, considering also that the absolute energy scale of the light pulses is not an important information, as long as the light emitted by different particles permits their discrimination.

\section{Alpha Background Rejection}
\label{sec:rejection}
\begin{figure}[thb]
\begin{centering}
\includegraphics[width=\columnwidth]{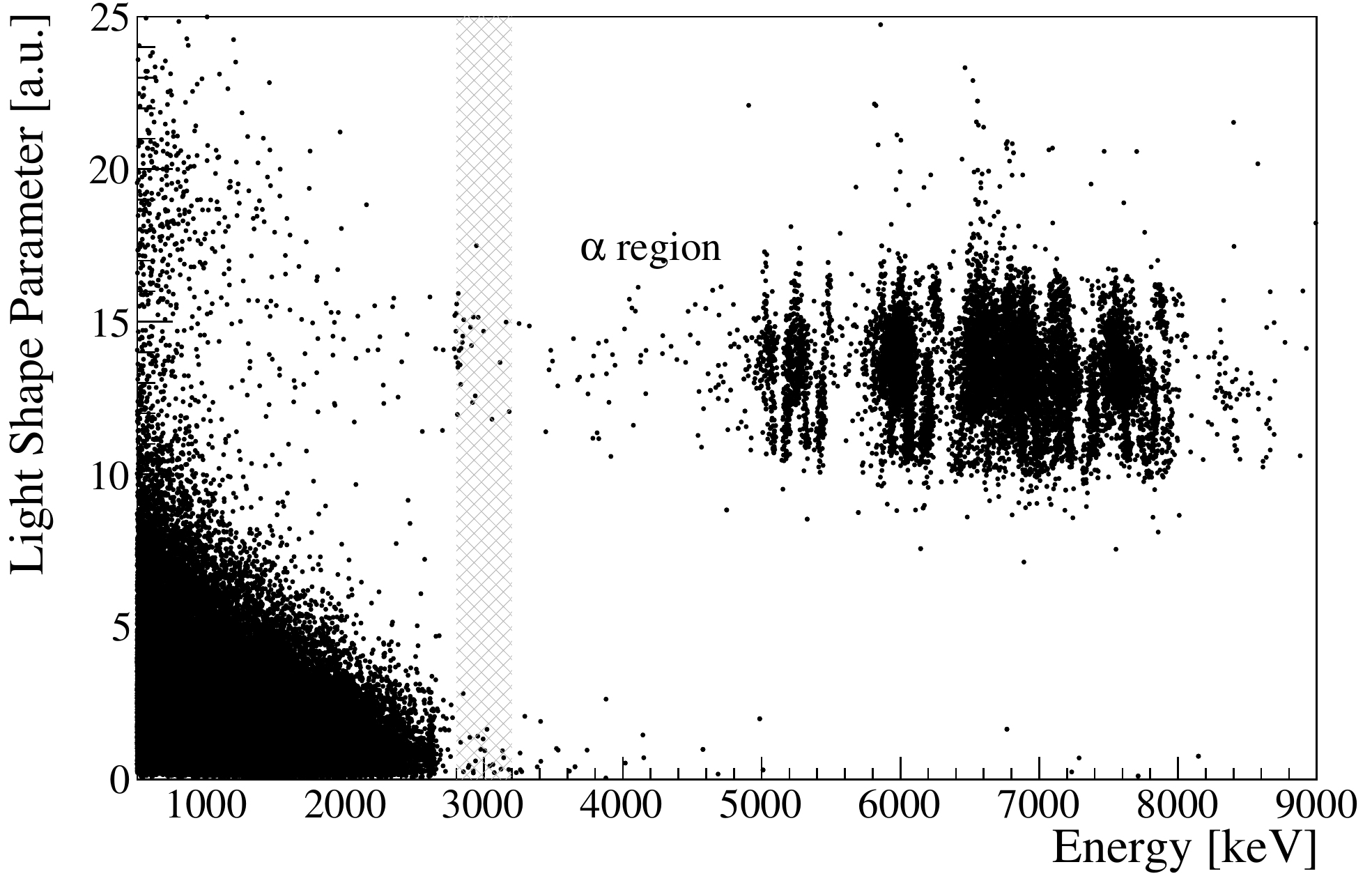}
\caption{Light Shape Parameter (SP) as a function of the heat released in the ZnSe (all crystals). The dashed region identifies the 400\,keV region centered around the \Se\ Q-value that is used for the \DBD\ analysis and for the estimation of the background.}
\label{fig:rejection}
\end{centering}
\end{figure}

The simultaneous read-out of the heat and light emitted by the ZnSe allows to reduce the background in the energy region of interest without affecting the signal efficiency.
After the selection of ``good'' thermal pulses (Sec.~\ref{sec:HeatCut}), we add the information provided by the LD.
We make a first selection by requiring each pulse to be associated to a non-zero light emission, to discard events that interacted in the temperature sensor, or electronics spikes and other spurious events not rejected by the (not aggressive) pulse-shape cuts.
Then, we study the shape of the light pulse $SP$ as a function of the heat released in the ZnSe crystal (Fig.~\ref{fig:rejection}). 

From a qualitative point of view, it is clear that the population of $\alpha$ events, that would produce a background of about 2$\times$10$^{-2}$ \ckky\ in the region of interest, can be clearly distinguished and rejected. However, the absence of peaks close to the \Se\ Q-value, as well as the energy-dependency of the $SP$, prevents a simple estimation of the signal efficiency and of the efficiency for the background rejection.

To compute the signal efficiency, we select a pure sample of $\beta/\gamma$ events in the ZnSe detectors: the ones that come from the electromagnetic showers produced by muons interacting in the materials that surround the detector. These events are produced in cascades, resulting in simultaneous triggers in several ZnSe crystals. For this reason, we select events in which at least five detectors triggered, obtaining a sample of 113 events. The sample is further selected by imposing a reasonable value for the detected light (larger than the noise fluctuations of the LD and smaller than the maximum light emitted by $\alpha$ particles). Indeed, a muon could cross the light detector and ionize it before producing the $\gamma$ cascade, and the effect of the ionization would be an unpredictable value of the $SP$. Finally, we remove events in which the light detectors feature more than one pulse in the same acquisition window that, again, could lead to a wrong calculation of the $SP$. 
\begin{figure}[thb]
\begin{centering}
\includegraphics[width=\columnwidth]{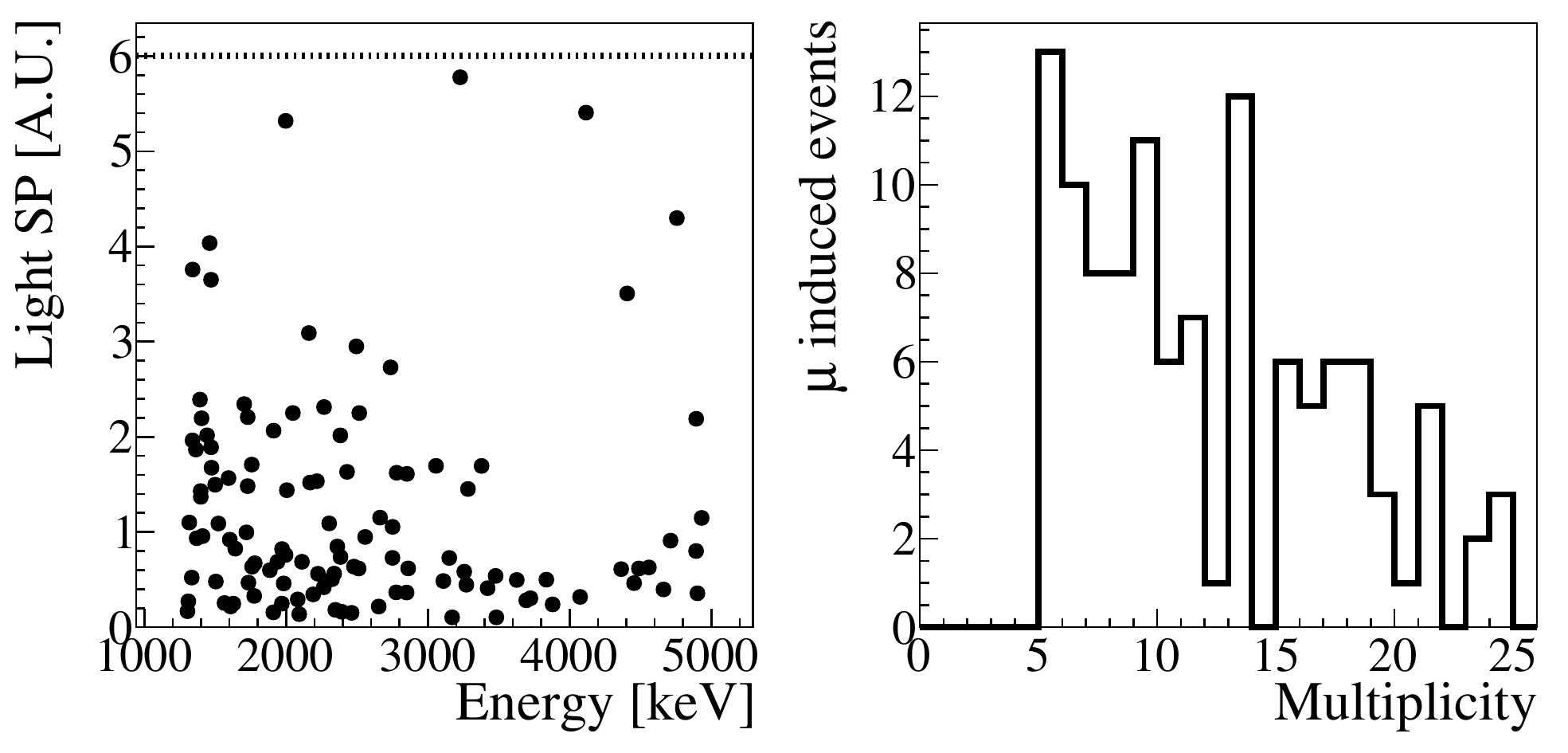}
\caption{Left: Light Shape Parameter distribution for muon-induced events as a function of the heat released in the ZnSe. The dotted black line represent the cut on SP. Right: histogram of the number of ZnSe crystals with a simultaneous trigger.}
\label{fig:muon}
\end{centering}
\end{figure}
Fig~\ref{fig:muon} shows the distribution of the $SP$ of the selected events which follows, as expected, the distribution of the electrons, and thus can be considered a good sample for the signal.
Looking at this distribution we set the cut $SP<6$, as this is the smallest value that yields a 100$\%$ efficiency on the signal.

This cut allows to reduce the background in the analysis region to \BetaGammaBkg\ \ckky.

Finally, we study the probability of mis-identifying an $\alpha$ interaction by selecting events with energy between 4 and 8\,MeV. 
The distribution of the $SP$ of these events can be modeled using a Gaussian function with mean value $\mu=$13.33$\pm$0.01 and $\sigma=1.38\pm0.01$.  
The probability for events that follow this distribution to occur below $SP=6$ (the selected cut) is 5$\times$10$^{-8}$, proving that the probability of a mis-identification is negligible. Even if it provides a satisfactory description of the data, the choice of modelling this distribution with a Gaussian function is not supported by physics considerations and could therefore lead to an underestimation of the background events in the region of interest, due to the presence of some outliers. Nevertheless, the number of events far from the cluster of alpha events is very small, proving that the large majority of the background events can be efficiently rejected.

Summarizing, the combination of light and heat allows to suppress the $\alpha$ background by almost a factor three without affecting the signal efficiency. Moreover, the alpha rejection capability already matches the requirements of next generation experiments, such as CUPID, in which the background must be close to zero at the tonne-scale level.

\section{Improving the Time Veto}
\label{sec:AlphaDelayed}
The identification of $\alpha$ particles down to low energies can help in reducing also the $\beta$/$\gamma$  background.
Indeed, one of the most worrisome background sources in the region of interest is $^{208}$Tl, an isotope belonging to the $^{232}$Th chain that decays via $\beta$/$\gamma$ with a Q-value of about 5\,MeV. 
Nevertheless, the $\beta$/$\gamma$ interactions produced by $^{208}$Tl can be efficiently rejected by exploiting the time coincidence with its parent,  $^{212}$Bi. 
This isotope decays to $^{208}$Tl with the emission of an $\alpha$ particle (Q-value of $\sim$6207\,keV), and  $^{208}$Tl subsequently decays with a half-life of about 3.05 minutes. Therefore, the $\beta$/$\gamma$ background from $^{208}$Tl can be suppressed by vetoing the detectors for a few minutes after the occurrence of an $\alpha$-like event with an energy corresponding to the $^{212}$Bi Q-value.

\begin{figure}[thb]
\begin{centering}
\includegraphics[width=\columnwidth]{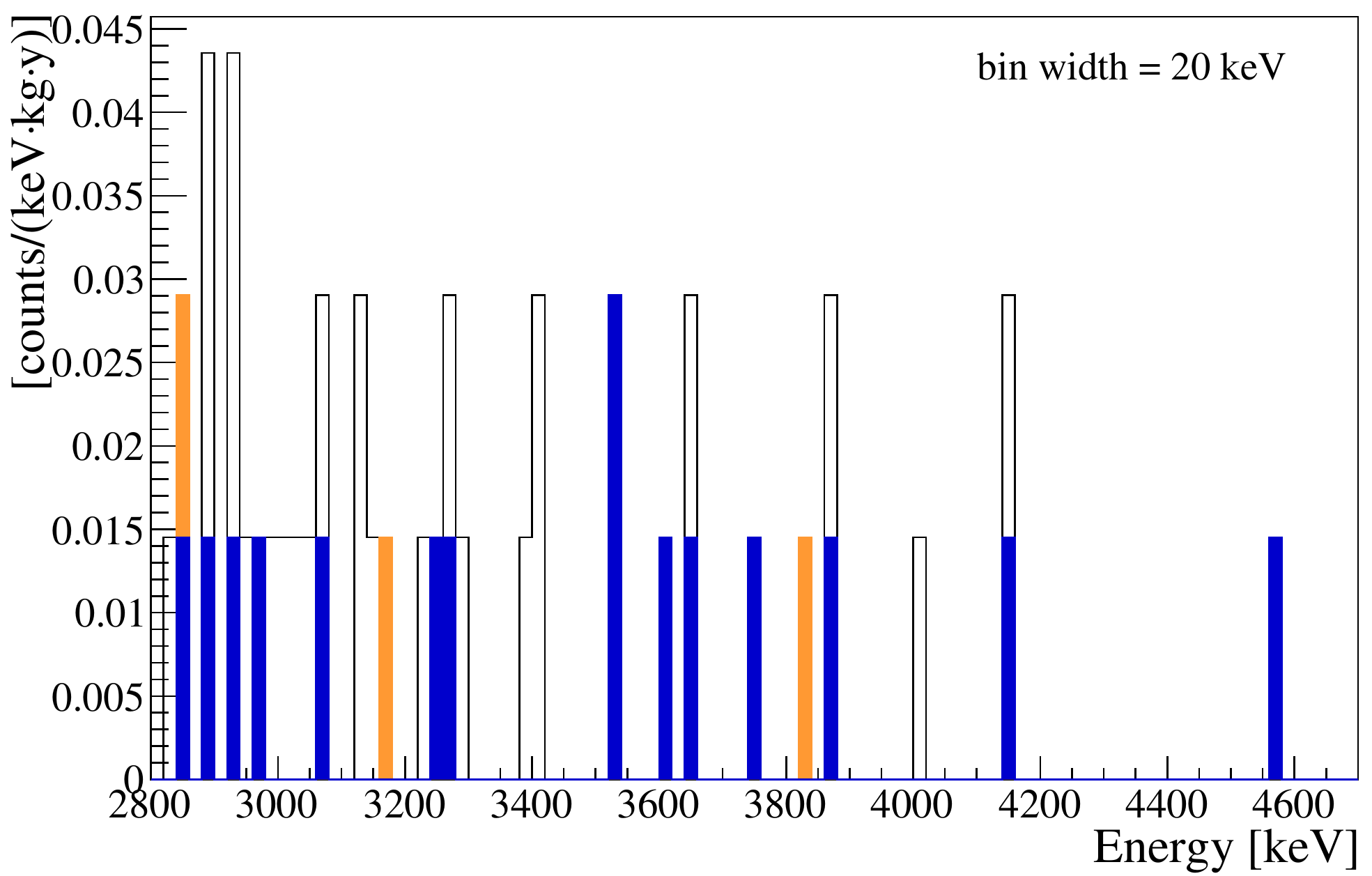}
\caption{Open histogram:  high-energy $\beta$/$\gamma$ spectrum of \CupidZ\ obtained with a ZnSe exposure of \exposure. Orange: the same spectrum after applying a time-veto of 3 half-lives after the detection of an $\alpha$ particle with energy compatible with the Q-value of $^{212}$Bi. Blue: events that survive a time-veto of 3 half-lives after the detection of an $\alpha$ particle with energy larger than 2\,MeV.}
\label{fig:delayed_coincidences}
\end{centering}
\end{figure}

This technique was already exploited in the past with satisfying results~\cite{Beeman:2013vda,Beeman:2011bg}.
To show its effect in \CupidZ, we report in Fig.~\ref{fig:delayed_coincidences} the high energy region of the $\beta/\gamma$ spectrum. 
This spectrum is obtained applying cuts on the pulse-shape, on the number of triggering ZnSe crystals (Sec.~\ref{sec:HeatCut}), and the $\alpha$ particles rejection (Sec.~\ref{sec:rejection}) and, as explained in the previous sections, results in \BetaGammaBkg \ckky.

From a Monte Carlo simulation taking as input the crystals contaminations (the reader can find in Ref.~\cite{Azzolini:2018tum} the $\alpha$ spectrum of the detector and the activities of the main peaks), we expect a large fraction of this background ((1.10$\pm$0.2)$\times$10$^{-2}$\ \ckky)  to be dominated by $^{208}$Tl.

As shown in Fig.~\ref{fig:delayed_coincidences}, the time-veto is very effective in the  abatement of high-energy $\beta/\gamma$ events. By applying a veto of 3 half-lives (3$\times$3.05\,min) after the detection of an $\alpha$ particle at the Q-value of $^{212}$Bi, the background reaches a value of \VetoedBkg\ \ckky\ with a dead-time of 1$\%$.

The innovative idea in \CupidZ\ consists in enlarging the window for the identification of an $\alpha$ produced by $^{212}$Bi down to much lower energies, by exploiting the excellent discrimination capability between $\alpha$'s and electrons. When the $^{212}$Bi decay occurs inside the crystal, indeed, it releases the whole decay energy ($\alpha$ + nuclear recoil) inside the detector, producing a characteristic peak at the Q-value of the transition. On the contrary, when the decay occurs on the crystal surface, or on the surface of the materials surrounding the crystal, the $\alpha$ particle can loose a variable fraction of its initial energy, resulting in a low-energy deposit inside the detector. 
By exploiting the possibility of disentangling $\alpha$ particles from electrons through the read-out of the scintillation light, we can tag also $^{212}$Bi interactions that do not produce a peak at the Q-value of the decay.
For this purpose, we select the possible $^{212}$Bi parents by choosing events with energy larger than 2\,MeV and light $SP$ between 7 and 25 (Fig.~\ref{fig:rejection}).

In Fig.~\ref{fig:delayed_coincidences} we compare the background obtained with a veto that exploits only the $^{212}$Bi peak at the Q-value, with the background obtained with a veto that exploits all the $\alpha$'s down to low energies.
In the latter case, the background in the region of interest becomes \FullVetoedBkg\ \ckky\ with a dead-time of 2.6$\%$.

\section{Summary and Conclusions}
In this paper we presented the analysis methods to exploit the simultaneous read-out of light and heat to suppress the background of cryogenic calorimeters. We showed a new method to create the signal templates for the matched filter, that allows to simplify the data processing and, at the same time, to select pulses as similar as possible to the expected signal.  We presented a technique to discriminate against $\alpha$ particles and a new method to estimate the efficiency on the signal and on the background rejection. Finally, we discussed a new time-veto for the suppression of the $^{208}$Tl background, that can deal with both internal and surface contaminations.
We summarize the main results of the analysis in Table~\ref{Table:background summary}.
\begin{table}[!htb]
\centering
\begin{tabular}{lcc}
\hline
Event Selection 			&Background Index 			&Efficiency\\
                          			&[\ckky\ ]                     &[$\%$]\\
\hline
Heat     					 &\TotalBkg  		                 &\CutEfficiency        \\
\hline
Heat + $\alpha$ rejection      	 &\BetaGammaBkg  	         	&\CutEfficiency       \\
\hline
Heat + veto with $^{212}$Bi       &\VetoedBkg   			         &94.5$\pm$2\%       \\
\hline
Heat + veto with all $\alpha$'s    &\FullVetoedBkg 		         &\TotEfficiency          \\
\hline
Total signal efficiency                            &\FullVetoedBkg 		         &75$\pm$2\%          \\
\hline
\end{tabular}
\caption{Summary of the Background Index (counts/keV/kg/y) and signal efficiency averaged on the \dataset s exposure, measured in the region 2800 -- 3200\, keV with a ZnSe exposure of {\exposure} ($1.34\times10^{25}$~emitters$\cdot$y). Uncertainties are reported at 68$\%$ C.L.. First row: events that pass the cuts on the heat described in Sec.~\ref{sec:HeatCut}. 
Second row: the events are further selected requiring that the shape parameter of the light is consistent with interactions of electrons ($\alpha$ rejection) as described in Sec.~\ref{sec:rejection}.
Third row: we added a time-veto of 3 half-lives after the detection of an $\alpha$ particle with energy compatible with the Q-value of $^{212}$Bi. 
Fourth row: we added a time-veto of 3 half-lives after the detection of an $\alpha$ particle with energy larger than 2\,MeV.
Last row: we report the total efficiency, including the data selection efficiency computed as in fourth row, the trigger efficiency, and the electrons containment efficiency of ($81.0\pm0.2$)~\% (see Ref~\cite{Azzolini:2018dyb}).}
\label{Table:background summary}
\end{table}

Thanks to the analysis tools presented in this paper we were able to prove that the simultaneous read-out of light and heat allows to reach a background of \FullVetoedBkg\ \ckky .
The achievement of this background level, the lowest among detectors based on cryogenic calorimeters, sets a key milestone for next generation experiments.

\begin{acknowledgements}
This work was partially supported by the LUCIFER experiment, funded by ERC under the European Union's Seventh Framework Programme (FP7/2007-2013)/ERC grant agreement n. 247115, funded within the ASPERA 2nd Common Call for R\&D Activities. We are particularly grateful to M.~Iannone for his help in all the stages of the detector construction,  A.~Pelosi for constructing the assembly line, M. Guetti for the assistance in the cryogenic operations, R. Gaigher for the mechanics of the calibration system, M. Lindozzi for the cryostat monitoring system, M. Perego for his invaluable help in several fields, the mechanical workshop of LNGS (E. Tatananni, A. Rotilio, A. Corsi, and B. Romualdi) for the continuous help in the overall set-up design. A.~S.~Zolotarova is supported by the IDI 2015 project funded by the IDEX Paris-Saclay, ANR-11-IDEX-0003-0.
This work makes use of the DIANA data analysis and APOLLO data acquisition software which has been developed by the Cuoricino, CUORE, LUCIFER and CUPID-0 collaborations.
\end{acknowledgements}

\bibliographystyle{spphys}       

\end{document}